\documentclass[aps,final,notitlepage,oneside,twocolumn,nobibnotes,nofootinbib,
superscriptaddress,noshowpacs,centertags]{revtex4-1}

\usepackage[english]{babel}
\usepackage{graphicx}
\usepackage{latexsym}
\usepackage{amssymb}
\usepackage{amsmath}
\usepackage{float}

\begin{document}

\title{Vaidya Spacetime in the Diagonal Coordinates}
\author{V.A. Berezin}\thanks{e-mail: berezin@inr.ac.ru}
\affiliation{Institute for Nuclear Research, Russian Academy of Sciences, 
pr. 60-letiya Oktyabrya 7a, Moscow, 117312 Russia}
\author{V.I. Dokuchaev}\thanks{e-mail: dokuchaev@inr.ac.ru}
\affiliation{Institute for Nuclear Research, Russian Academy of Sciences, 
pr. 60-letiya Oktyabrya 7a, Moscow, 117312 Russia}
\author{Yu.N. Eroshenko}\thanks{e-mail: eroshenko@inr.ac.ru}
\affiliation{Institute for Nuclear Research, Russian Academy of Sciences, 
pr. 60-letiya Oktyabrya 7a, Moscow, 117312 Russia}

\date{\today}

\begin{abstract}
We have analyzed the transformation from initial coordinates $(v,r)$ of the Vaidya metric with light 
coordinate $v$ to the most physical diagonal coordinates $(t,r)$. An exact solution has been obtained for the corresponding metric tensor in the case of a linear dependence of the mass function of the Vaidya metric on light 
coordinate $v$. In the diagonal coordinates, a narrow region (with a width proportional to the mass growth rate 
of a black hole) has been detected near the visibility horizon of the Vaidya accreting black hole, in which the 
metric differs qualitatively from the Schwarzschild metric and cannot be represented as a small perturbation. 
It has been shown that, in this case, a single set of diagonal coordinates $(t,r)$ is insufficient to cover the entire 
range of initial coordinates $(v,r)$ outside the visibility horizon; at least three sets of diagonal coordinates are 
required, the domains of which are separated by singular surfaces on which the metric components have singularities 
(either $g_{00}=0$ or $g_{00}=\infty$.). The energy-momentum tensor diverges on these surfaces; however, the 
tidal forces turn out to be finite, which follows from an analysis of the deviation equations for geodesics. 
Therefore, these singular surfaces are exclusively coordinate singularities that can be referred to as false firewalls 
because there are no physical singularities on them. We have also considered the transformation from 
the initial coordinates to other diagonal coordinates $(\eta,y)$, in which the solution is obtained in explicit form, 
and there is no energy-momentum tensor divergence.
\end{abstract}

\maketitle 




\section{INTRODUCTION}

The Vaidya metric describes the spacetime produced
by a spherically symmetric radial radiation flow.
This metric has the form \cite{Vaidya43,Vaidya51,Vaidya53}
\begin{equation}
ds^2=\left[1-\frac{2m(z)}{r}\right]dz^2+2dzdr-r^2(d\theta^2+\sin^2\theta d\varphi^2).
\label{Vaidya}
\end{equation}
In particular, the Vaidya metric describes a nonstationary
accreting or emitting black hole. In this metric,
$m(z)$ is an arbitrary mass function that depends
(in the case of accretion) on coordinate $z=-v$, where
$v$ is the advanced light coordinate or (in the case of
emission of radiation) on coordinate $z=u$, where $u$ is
the retarded light coordinate. For $m(z)=m_0=const$,
metric (\ref{Vaidya}) describes a Schwarzschild black hole of
mass $m=m_0$. Here and below, we are using units of
measurement in which $c=1$ for the velocity of light
and $G=1$ for the gravitational constant.

Vaidya metric (\ref{Vaidya}), which is one of a few known
exact solutions in the general theory of relativity, has a
large number of astrophysical and theoretical applications. In particular, it is used to describe the quantum
evaporation of black holes \cite{VolZagFro76,Hiscock81,Kuroda84,Beciu84,Kaminaga90,Zheng94,Farley06,Sawayama06} or the emission of
radiation by astrophysical objects \cite{Knutsen84,Barreto93,Adams94,Maharaj02,Sungwook10,Alishahiha14}. This metric
is also employed in investigations of gravitational collapse
and the formation of naked singularities \cite{HiscockWilliamsEardley,Kuroda84b,Papapetrou85,WauLak86b,Dwivedi89,Dwivedi91,Joshi92, Joshi92b,Joshi93,Dwivedi95,Ghosh01,Mkenyeleye14}. However, the interpretation of physical results
obtained in this metric is complicated because this
metric is written in terms of coordinates $(z,r)$, where $z$
is not a directly measurable physical quantity. It is
known that, in some simple cases, the expressions that
describe the transition to double zero coordinates
$(v,u)$ can be derived \cite{WauLak86}, but a transition to more
physical diagonal coordinates involves analytic difficulties,
and the explicit form of the corresponding
coordinate transformation is generally unknown \cite{LinSchMis65}.

In this study, we analyze the coordinate transformations
from the standard coordinates $(z,r)$ of the
Vaidya metric to diagonal coordinates in the case of a
linear mass function $m(z)=-\alpha z+m_0$, $dm/dz=-\alpha=const$, where parameter $\alpha>0$ characterizes the accretion
or emission rate. Using this Ansatz, we solve the
problem of transforming the Vaidya metric to the diagonal
coordinates fully analytically by calculating all
metric coefficients. Vaidya metric (\ref{Vaidya}) with a linear
mass function has been considered previously in a
large number of publications in various aspects \cite{VolZagFro76,Hiscock81,LevinOri,Shao05,AbdallaChirenti,YangJeng06,BengtssonSenovilla09}. However, the form of the Vaidya metric in
diagonal coordinates was not obtained in these publications.
We obtained the first such solution in \cite{we},
where special diagonal coordinates $(\eta,y)$ were used
(these coordinates will be considered in Section~\ref{etaycoordsec})
below). In this work, we will also obtain the solution
for another (more physical) choice of diagonal coordinates
$(t,r)$.

The transition to the diagonal coordinates makes
the Vaidya problem closer to the actual situation
because these coordinates correspond to the results of
physical measurements that could have been taken by
a static observer. Using the diagonal coordinates, it is
clear how the accretion process is seen by the static
observer of a black hole or, in a more general form, the
physical structure of space-time in the presence of a
radial radiant flux. Therefore, this formulation of the
problem is extremely close to the physically realizable
situation.

It turned out that, even in the region beyond the
gravitational radius $r>2m$, in the $R$-region, a single set
of diagonal coordinates $(t,r)$ or $(\eta,y)$ is insufficient to
cover the entire range of variations in initial coordinates
$(v,r)$ in metric (\ref{Vaidya}), but several sets of diagonal
coordinates are required, the ranges of variations in
which are separated by surfaces with singularities of
the metric (either $g_{00}=0$ or $g_{00}=\infty$). These sets served
as the charts that cover the entire manifold with allowance
for physical limitation $m\geq0$. At the boundaries of
these charts, the energy-momentum tensor experiences
a divergence, which, however, is not associated
with the presence of physical caustics in the distribution
of accreted radiation. Analysis of the deviation
equations of geodesics shows that tidal forces on the
boundary surfaces are finite; therefore, these surfaces
are physically coordinate singularities that can be
referred to as false firewalls.

Initial Vaidya metric (\ref{Vaidya}) is geodetically incomplete
and requires analytic expansion for describing the
global geometry of space-time. One of the expansions
was proposed by Izrael \cite{Israel67} in general form for the
global geometry of eternal space-time with infinite ladders
of black and white holes. Other approaches were
used in \cite{Kuroda84,Fayosyx95}, in which additional spacetime regions
were constructed, as well as in \cite{HiscockWilliamsEardley,Krori74,Waugh86}, where special
mass fractions were employed. The application of
new diagonal coordinates enabled us to reveal the
global structure of the space-time for the Vaidya metric
with linear mass function $m(z)$. The main instruments
of analysis are exact expressions for the radial light
geodesics. As a result, we have constructed a geodetically
complete (with physical limitation $m\geq0$) space-time
and the corresponding conformal Carter--Penrose diagrams.

\section{VAIDYA METRIC IN DIAGONAL COORDINATES $(t,r)$}
\label{trsec}

In this section, we choose coordinates of curvatures
$(t,r,\theta,\phi)$ as diagonal coordinates, in which the
metric has the form \cite{LL-2}
\begin{equation}
 \label{spherical2}
 ds^2=e^{\nu(t,r)} dt^2-e^{\lambda(t,r)}dr^2-r^2(d\theta^2+\sin^2\theta\,d\phi^2)
 \end{equation}
or, after the redefinition of the coefficients for convenience,
\begin{equation}
 ds^2=f_0(t,r) dt^2-\frac{dr^2}{f_1(t,r)}-r^2\left(d\theta^2+\sin^2\theta d\phi^2\right),
\label{spherical-ini}
\end{equation}
where $f_0(t,r)$ and $f_1(t,r)$ are certain functions that can
be determined from the Einstein equations. Let us
introduce mass function $M_1(t,r)$ connected by definition
with $\lambda$ and $f_1(t,r)$ by the following relation:
\begin{equation}
 e^{-\lambda(t,r)}= f_1(t,r)=1-\frac{2M_1(t,r)}{r}.
\label{m1n}
\end{equation}


\subsection{Transition to Diagonal Coordinates}
   \label{diagtranssub}
   
We will seek the transformation of coordinates of
the initial Vaidya metric to diagonal coordinates as follows:
\begin{equation}
z=z(t,\tilde{r}), \quad r=\tilde{r}.
\label{transf}
\end{equation}
Substituting these relations into (\ref{Vaidya}) and equating the
resulting coefficients to the corresponding coefficients
in Eq. ~(\ref{spherical-ini}), we obtain the system of equations
\begin{equation}
f_0=f_1\dot{z}^2, \quad z'=-\frac{1}{f_1}.
\label{f0f1zsh}
\end{equation}
We write the second of these equations in the form
\begin{equation}
z'=-\frac{1}{1-\frac{2m(z)}{r}}
\end{equation}
and multiply it by $dm/dz$:
\begin{equation}
\frac{dm}{dz}z'=M_1'=-\frac{\frac{dm}{dz}}{1-\frac{2m(z)}{r}}.
\label{dmdz}
\end{equation}

We consider the linear mass function
\begin{equation}
m(z)=-\alpha z+m_0.
\label{linanz}
\end{equation}
In the case of accretion, we have $z=-v$ and $m=\alpha v+m_0$. In the case of emission, we obtain $z=u$ and $m=-\alpha u+m_0$. Consequently, for both cases, $\alpha>0$ and
$dm/dz=-\alpha$. Then, Eq.~(\ref{dmdz}) assumes the form
\begin{equation}
M_1'=\frac{\alpha}{1-\frac{2M_1}{r}}.
\label{dmdz2}
\end{equation}
Denoting
\begin{equation}
y=1-\frac{2M_1(t,r)}{r},
\end{equation}
we write the solution to Eq.~(\ref{dmdz2}) in the following
implicit form:\\
\begin{equation}
-\int\frac{ydy}{y^2-y+2\alpha}=\ln\frac{r}{r_0}+\phi(t).
\label{intsol}
\end{equation}

Let us first consider the case when $\alpha<1/8$. Evaluating
the integral in Eq.~(\ref{intsol}), we obtain
\begin{equation}
\frac{r}{r_0}B(t)=\Psi(y),
\label{rsol}
\end{equation}
where the following notation has been introduced:
\begin{equation}
\Psi(y)=|y-y_1|^{\frac{y_1}{y_2-y_1}}|y-y_2|^{-\frac{y_2}{y_2-y_1}}
\label{phiexpr}
\end{equation}
and
\begin{equation}
y_1=\frac{1-\sqrt{1-8\alpha}}{2}, \qquad y_2=\frac{1+\sqrt{1-8\alpha}}{2},
\label{y12}
\end{equation}
and $B(t)=e^{\phi(t)}>0$ is a certain as yet unknown function.

Let us now find coefficient $f_0$. Differentiating
Eq.~(\ref{rsol}) with respect to $t$ and substituting
\begin{equation}
\dot{y}=-\frac{2\dot M_1}{r}=\frac{2\alpha \dot{z}}{r},
\end{equation}
we obtain
\begin{equation}
\frac{r^2}{2\alpha  r_0^2}=\frac{d\Psi}{dy}\frac{\dot{z}}{\dot{B(t)}}.
\end{equation}
Function $\Psi(y)$ has singular points $y=y_1$ and $y=y_2$;
therefore, the entire domain $-\infty<y<1$ splits into four
parts as follows:
\begin{equation}
-\infty<y<0, \quad 0<y<y_1, \quad y_1<y<y_2, \quad y_2<y<1,
\label{yregions}
\end{equation}
in each of which we must carry out separate calculations.
It should be noted that free parameters $r_0$ and $t_0$
can be different in different regions, and the relation
between them must be established by joining the solution.
The time can be redefined as follows: $d\tilde{t}^2=\dot{B}^2dt^2$; therefore, we can choose
\begin{equation}
\dot{B}=\pm\alpha, \quad B(t)=\pm\alpha(t-t_0),
\end{equation}
where the sign in each specific case is determined from
the condition $\dot{z}>0$ as follows:
\begin{equation}
B(t)=\left\{
  \begin{tabular}{ccc}
    $\alpha (t-t_0)$ & ~~ & $y<0$,\\
    $-\alpha (t-t_0)$ & ~~ & $0<y<y_1$,\\
     $\alpha (t-t_0)$ & ~~ & $y_1<y<y_2$,\\
    $-\alpha (t-t_0)$ & ~~ & $y>y_2$.
  \end{tabular}
\right.
\label{c1sol3}
\end{equation}
In all domains (\ref{yregions}), we obtain the same expression
from Eqs.~(\ref{f0f1zsh}) as follows:
\begin{equation}
f_0=y\dot{z}^2=\frac{1}{y}|y-y_1|^{\frac{2y_2}{y_2-y_1}}|y-y_2|^{-\frac{2y_1}{y_2-y_1}}.
\label{f0expr}
\end{equation}
Thus, we have determined all metric coefficients in
parametric form as (\ref{f0expr}), $f_1=y$, and (\ref{rsol}), where $y$ is a
parameter.

\begin{figure}[t]
\begin{center}
\includegraphics[angle=0,width=0.49\textwidth]{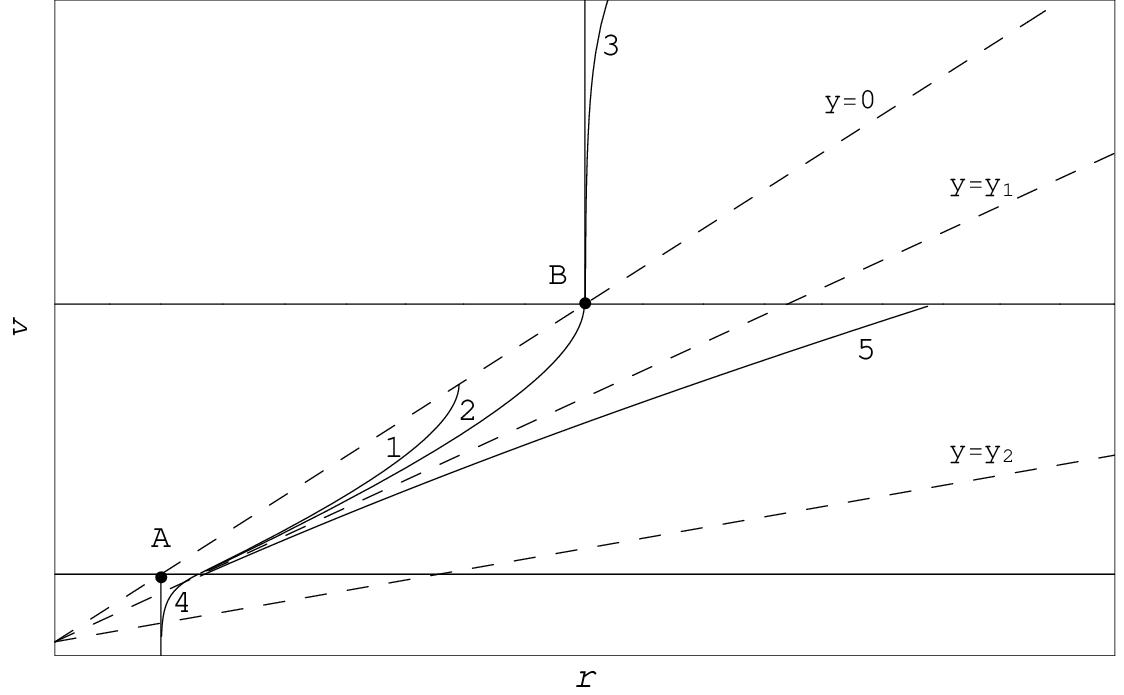}
\end{center}
\caption{Different branches of solution (\ref{rsol}) are shown by
solid lines 1, 2, and 5 (lines of constant $t$). Dashed lines
separate different ranges of variation of parameter $y$. It is
assumed that Eq.~(\ref{linanz}) is valid above point A and below
point B; in other regions, $M_1=const$. This means that a
spherical layer (thick shell) of radiation enclosed between
two horizontal lines passing through A and B is incident
on the black hole. Therefore, the visibility horizon is
shown by the vertical ray passing downwards from point
A, segment AB, and the vertical ray passing upwards from
point B. Below A and above B, Schwarzschild solution
(\ref{funvsch}) is valid, but with different constants $M_1$. By way of
example, these solutions for the Schwarzschild metric are
shown by curves 3 and 4 joined with solution 2 in the
region inside the shell. At point B, the solution suffers
discontinuity: particles move under the visibility horizon
over finite time $t$ below this point, while above this point,
for $t\to\infty$, photons approach the visibility horizon, but
not attain it. Line 2 is the last line of constant time before
the discontinuity of the solution, while line 1 is the line of
constant time for an earlier instant $t$.}
\label{gr1}
\end{figure}

The differential transformation to new coordinates
has the form
\begin{equation}
dz(t,r)=f_0^{1/2}f_1^{-1/2}dt-f_1^{-1}dr.
 \label{dudtdr}
\end{equation}
It was noted in \cite{LinSchMis65} that the expression for the differential
analogous to (\ref{dudtdr}) could have been obtained by
evaluating the integrating factor. It is important that
$M_1$ in the new metric is a function of $z$ only; i.e.,
$M_1(t,r)=M_1(z)$, where $z=z(t,r)$. Figure~\ref{gr1} shows the
sections of function $z=-v(t,r)$ (in the case of accretion)
by planes$t=const$.

Let us consider the limiting transition to the
Schwarzschild metric for $\alpha\to0$. In Fig.~\ref{gr1}, the
Schwarzschild metric is above point B; therefore,
asymptotically flat infinity is outside the thick shell.
Within the shell, between points A and B, we have
$y_1\to2\alpha\to0$, $y_2\to1-2\alpha$, $-2\beta_2\to2$, and $-2\beta_2\to-4\alpha$ for $\alpha\to0$; hence, expression (\ref{f0expr}) has the asymptotic
form (it should be recalled that $y=f_1$) as follows:
\begin{equation}
f_0\to f_1\left|\frac{2M_1}{r}+2\alpha\right|^{-4\alpha}.
\end{equation}
At any radius $r=const$ for $\alpha\to0$, we obtain $|...|^{-4\alpha}\to1$ and, ultimately, $f_0\to f_1$, i.e., the Schwarzschild metric.
The deviation of the Schwarzschild metric is only observed in a narrow region of width approximately
equal to $\sim\alpha$ near the gravitational radius.


\subsection{Light Beams in the Diagonal Metric}

Let us consider the propagation of light beams in
the Vaidya metric corresponding to accretion ($z=-v$). The behavior of the emergent zeroth geodesic
determines the event horizon. For linear function
$M_1(z)$, light geodesics in the $(v,r)$ coordinates were
analyzed in \cite{VolZagFro76}. From relation $ds^2=0$ in the Vaidya
metric (\ref{Vaidya}), we obtain the following equation for the
emergent beam:
\begin{equation}
2rdr=[r-2M_1(v)]dv.
\label{ish}
\end{equation}
Using Eq.~(\ref{dudtdr}), we can easily show that this equation
in the $(t,r)$ coordinates has the form $dt=dr/\sqrt{f_0f_1}$.
The structure of solutions to Eq.~(\ref{ish}) for function $r(t)$
is analogous to the structure of above solutions (\ref{rsol})
and (\ref{phiexpr}) for sections $t=const$. The equations of
motion of the emergent light beam in parametric form
can be written as
\begin{eqnarray}
t(y)&=&\int\frac{dr}{\sqrt{f_0f_1}},
\label{sol2t}
\\
\nonumber
\\
r(y)&=&|y-y_3|^{\frac{y_3}{y_4-y_3}}|y-y_4|^{\frac{-y_4}{y_4-y_3}}D,
\label{sol2}
\end{eqnarray}
where
\begin{equation}
y_3=\frac{1-\sqrt{1-16\alpha}}{2}, \qquad y_4=\frac{1+\sqrt{1-16\alpha}}{2},
\label{z12}
\end{equation}
parameter $y=f_1$ is the same as before, and constant
$D>0$ labels the beams. Integral (\ref{sol2t}) has a rather cumbersome
form and cannot be evaluated analytically;
however, the exact form of function $t(y)$ will not be
required for further analysis. There is a separatrix that
divides the solutions bounded in the radius from
unfounded solutions. This means that, for each radius,
there is an instant such that a photon emitted prior to
this instant can go to infinity. However, if a photon is
emitted later, it only reaches a finite radius. This
behavior of light geodesics will be clarified in
Section~\ref{etaycoordsec}, where the meaning of surfaces $y=y_3$ and
$y=y_4$ will be considered.

If a linear approximation, $M_1(z)$ terminates at a
certain $z$, the visibility horizon in the $(z,r)$ coordinates
is the emergent light beam passing through the point
of joining of two regions (via point B in Fig.~\ref{gr1}). The
method of joining of different regions should be considered
separately.

Let us now consider the incident (propagating to
the center) beams and determine the time of flight of
a photon until it crosses the visibility horizon. It
should be noted that the event of crossing occurs
under the global event horizon and, hence, is inaccessible
for observation from outside the event horizon.
The incident radial beam obeys the equation $v=const$, which has the form $dt=-dr/\sqrt{f_0f_1}$ in the $(t,r)$
coordinates. Substituting $f_1=y$ and $f_0$ determined in
accordance with Eq.~(\ref{f0expr}), we obtain
\begin{eqnarray}
&&\Delta t=2M_1\int\limits_y^{y_i}\frac{dx|x-y_1|^{\frac{-y_2}{y_2-y_1}}|x-y_2|^{\frac{y_1}{y_2-y_1}}}{(1-x)^2}= \nonumber
\\
&=&M_1(1-x)^{-2}\times
\label{dtitog}
\\
&\times&\left.F_1\left[2;\frac{y_2}{y_2-y_1},-\frac{y_1}{y_2-y_1};3,\frac{1-y_1}{1-x},\frac{1-y_2}{1-x}\right]\right|_{x=y}^{x=y_i},\nonumber
\end{eqnarray}
where $F_1[...]$ is the Appell hypergeometric function
and $y_i$ and $y$ are the initial and final values of parameter
$y=1-2M_1/r$. In evaluating the integral in Eq.~(\ref{dtitog}),
we take into account the fact that we consider the
beam with $v=const$; therefore, $M_1(v)$ can be removed
from the integrand. It can be seen from relation (\ref{dtitog})
that the light beam reaches surface $y=y_2$ during finite
coordinate time $\Delta t$ and surface $y=y_1$ over infinite time
$\Delta t=\infty$. Consequently, the beam can cross the visibility
horizon $y=0$ if the initial point of its trajectory is
chosen for $y<y_1$. It should be noted that the partial
analog of $\Delta t\to\infty$ is the infinite time of attainment of
the gravitational radius of a Schwarzschild black hole
by test particles, which is determined by the singular
behavior of the Schwarzschild coordinates on the
event horizon. The case under investigation differs in
that the condition $\Delta t\to\infty$ is observed on surface $y=y_1$ outside the visibility horizon due to the influence of
accreting matter on the metric.

In the slow accretion limit $\alpha\to0$ (but $\alpha\neq0$), for
$y_i\ll y_1$, $y=0$, we obtain the time of flight up until the
visibility horizon is crossed
\begin{equation}
\Delta t\simeq-2M_1\ln\alpha.
\label{dtzero}
\end{equation}

In the case of limiting the transition to the
Schwarzschild metric with $\alpha=0$, the left branch of
solution (\ref{rsol}) (Fig.~\ref{gr1}, curves 1, 2) becomes degenerate,
and we must choose the right branch (curve 5) of the
solution for $y_1<y<y_2$. This is due to the fact that, as
noted above, accretion near the visibility horizon
changes the geometry qualitatively. If we perform a
limiting transition in this way (choose the right
branch), we find that $y_1\to0$, $y_2\to1$, $y_1/(y_2-y_1)\to0$, $y_2/(y_2-y_1)\to-1$, and expression (\ref{dtitog}) is transformed
into the exact expression for the time of flight of a
photon in the Schwarzschild metric.


   \subsection{Geometrical and Physical Meanings
of Surfaces $y=y_1$ and $y=y_2$}

To clarify the origin of lines $y=y_1$ and $y=y_2$, which
are absent in the Schwarzschild solution for $\alpha=0$, we
analyze the surfaces $y=const$. Let us first calculate the
square of the normal to these surfaces. Since it is an
invariant, this can be done in any metric (most easily,
in initial Vaidya metric (\ref{Vaidya})). We denote this invariant
by $Y$ as follows:
\begin{equation}
Y=\gamma^{ik}y_{,i}y_{,k}=\frac{(1-y)^3}{4m^2}(y-y_3)(y-y_4),
\label{Y0}
\end{equation}
where $y_3$ and $y_4$ are defined by formulas (\ref{z12}).

Let us calculate invariant $Y$ along the emergent
light beam, substituting expression $m=r(1-y)/2$ and
$r(y)$ from Eq.~(\ref{sol2}) into expression (\ref{Y0}) as follow:
\begin{eqnarray}
Y&=&\frac{(1-y)}{D^2}|y-y_3|^{\frac{y_4-3y_3}{y_4-y_3}}|y-y_4|^{\frac{3y_4-y_3}{y_4-y_3}}\times
\nonumber
\\
&\times&{\rm sign}(y-y_3){\rm sign}(y-y_4).
\label{Y0-1}
\end{eqnarray}
Above all, it can be seen that there are no singularities
on lines $y=y_1$ and $y=y_2$. The singularities observed
earlier on lines $y=y_1$ and $y=y_2$ are of purely coordinate
origin and correspond to the termination of operation
of the coordinate systems on lines $y=y_1$ and $y=y_2$. In fact, the action of individual systems of coordinates
terminates on these surfaces; these surfaces are
the boundaries of the coordinate charts covering the
entire manifold, while several sets of coordinates are
required to cover the entire spacetime in the diagonal
coordinates. In addition, we have $y_4-3y_3\geq0$ for $\alpha<3/64$ and $y_4-3y_3<0$ for $\alpha>3/64$ in the exponent,
while $3y_4-y_3>0$ in all cases. This means that invariant $Y$ changes its sign on lines $y=y_3$ and $y=y_4$ and
may vanish or turn to infinity. If we move along the
incident beam, when $m=const$, invariant (\ref{Y0}) vanishes
for $y=y_3$ and $y=y_4$. Nevertheless, lines $y=y_3$
and $y=y_4$ are not physical singularities of the metric
either, these lines are eliminable coordinate singularities
(this will be shown in Section~\ref{etaycoordsec}).

To clarify the physical meaning of surfaces $y=y_1$
and $y=y_2$, we calculate the energy-momentum tensor
using the resultant exact solutions. The first three Einstein
equations in metric (\ref{spherical2}) in the general case have
the form \cite{LL-2}
\begin{eqnarray}
 \label{G01}
 8\pi T^1_0 &=& -e^{-\lambda}\frac{\dot\lambda}{r}, \\
 \label{G00}
 8\pi T^0_0 &=& -e^{-\lambda}\left(\frac{1}{r^2}-
 \frac{\lambda'}{r}\right)+\frac{1}{r^2}, \\
 \label{G11}
 8\pi T^1_1 &=& -e^{-\lambda}\left(\frac{1}{r^2}+
 \frac{\nu'}{r}\right)+\frac{1}{r^2}. 
\end{eqnarray}
We denote $F\equiv\sqrt{f_0f_1}$. For the radial motion of photons
to the center, we can write $k^{\mu}=(a,b,0,0)$, $a>0$, $g_{\mu\nu}k^{\mu}k^{\nu}=f_0a^2-b^2/f_1=0$,  $k^{\mu}=a(1,-F,0,0)$, $k_{\mu}=a(f_0,F/f_1,0,0)$. The energy-momentum tensor has the form $T_\mu^\nu=\gamma k_\mu k^\nu$, $T_0^0=\gamma a^2f_0$, $T_0^1=-\gamma a^2f_0F$.
Expression (\ref{G01}) yields
\begin{equation}
\gamma a^2=\frac{\dot M_1}{4\pi r^2f_0F},
\end{equation}
then
\begin{equation}
T_0^0=-T_1^1=\frac{\dot M_1}{4\pi r^2F},~~~T_0^1=-\frac{\dot M_1}{4\pi r^2}.
\label{t00t01}
\end{equation}
The case when $\dot M_1>0$ corresponds to accretion and
$\dot M_1<0$ corresponds to emission. Substituting the expressions
into (\ref{G00}), we obtain
\begin{equation}
M_1'=\frac{\dot M_1}{F}=4\pi r^2T_0^0.
\label{dashdot0}
\end{equation}

Let us now consider the special case of linear
dependence $M_1(z)$. Substituting 
\begin{equation}
\dot{M_1}=\frac{dm}{dz}\dot{z}=-\alpha\frac{f_0^{1/2}}{f_1^{1/2}}
\label{dashdot}
\end{equation}
into (\ref{t00t01}) and raising the index, we obtain
\begin{equation}
T^{00}=\frac{\alpha}{4\pi r^2f_0f_1}, \quad T^{01}=-\frac{\alpha}{4\pi r^2f_0^{1/2}f_1^{1/2}}.
\end{equation}

Let us consider these expressions along the incident
light beam. In this case,$v=const$, $m=const$,
and, hence, $r^2=(2m)^2/(1-y)^2$ is a regular function for
$y=y_1$ and$y=y_2$. At the same time, the expression
obtained from (\ref{f0expr}) (it should be recall that $f_1=y$),
\begin{equation}
f_0f_1=|y-y_1|^{\frac{2y_2}{y_2-y_1}}|y-y_2|^{-\frac{2y_1}{y_2-y_1}}
\end{equation}
has singularities at points $y=y_1$ and $y=y_2$. The equation
for the light beam gives $f_0^{1/2}f_1^{1/2}=-dr/dt$; therefore,
we have
\begin{equation}
T^{00}=\frac{\alpha}{4\pi r^2(dr/dt)^2}, \quad T^{01}=\frac{\alpha}{4\pi r^2(dr/dt)}
\end{equation}
along the incident beam. Therefore, energy-momentum
tensor components $T^{00}$ and $T^{01}$ vanish for $y=y_2$
and turn to infinity for $y=y_1$ for the kinematic reason
(due to the existence of limiting points $dr/dt=0,\infty$ for
a light beam moving in coordinates $(t,r)$. For photons,
$y=y_1$ is the surface of infinitely large red shift
($g_{00}\to 0$); accretion has led to splitting of the former
Schwarzschild horizon, while $y=y_2$ is the surface of
infinitely large blue shift ($g_{00}\to\infty$); the emergence of
this surface distinguishes qualitatively the resultant
metric from the Schwarzschild metric. It is also
important to note that lines $y=y_1$ and $y=y_2$ are spatially
similar. It should be emphasized that the divergence
of $T^{00}$ and $T^{11}$ is not associated with the existence
of physical caustic and is of purely coordinate
origin. The operation of coordinate systems terminates
on lines $y=y_1$ and $y=y_2$, and singularities
appear in the behavior of the coordinates. These singularities
are also determined by the character of time
coordinate $t$ because radius $r$, which is an invariant,
does not experience changes on lines $y=y_1$ and $y=y_2$.
It is impossible to get rid of these singularities using
this coordinate system.

Let us analyze the above-mentioned coordinate
singularities using the deviation equations for geodesics
in the case of massive particles as follows:
\begin{equation}
\frac{D^2v^\mu}{ds^2}=R^\mu_{~\nu\rho\sigma}u^\nu u^\rho v^\sigma,
\label{devgeod}
\end{equation}
where $s$ is the interval chosen as an affine parameter,
$u^\mu$ is a vector tangent to a geodesic line, and $v^\mu$ is the
vector separating two geodesic lines. Let us choose a
purely spatial radial vector $v^\mu=(0,v,0,0)$ in diagonal
coordinate system (\ref{spherical-ini}); then, this vector in the Vaidya
metric has the form $v^\mu=(v/f_1,v,0,0)$. In both systems,
the radial component of vector $u^\mu$ is $u^1=u$. Let
us calculate the right-hand side of Eq.~(\ref{devgeod}) first in
Vaidya metric (\ref{Vaidya}), then transform it to a diagonal metric
(\ref{spherical-ini}). In the Vaidya metric, Eq.~(\ref{devgeod}) only contains
component $R_{0110}=2m/r^3$, while other components of
the curvature tensor contain angular indices and do
not appear in the result. Ultimately, we obtain the following
expression for the spatial part in the diagonal
coordinates:
\begin{equation}
R^1_{~\nu\rho\sigma}u^\nu u^\rho v^\sigma=\frac{2mu^2v}{r^3f_1}.
\label{devgeod2}
\end{equation}
Let us consider the behavior of the emergent geodesic
with $u<0$ in the Vaidya metric. For this, we write the
following equation for the radial component of the
geodesic:
\begin{equation}
\frac{du}{ds}+\Gamma^1_{00}(u^0)^2+\Gamma^1_{01}u^0u^1=0,
\label{georad}
\end{equation}
where
\begin{equation}
\Gamma^1_{00}=\frac{-r^2(dm/dz)-2m^2+rm}{r^3}, \quad \Gamma^1_{01}=-\frac{m}{r^2},
\end{equation}
and $u^0$ can be determined from the normalization condition
$u^\mu u_\mu=1$ in the form
\begin{equation}
u^0=\frac{u\pm\sqrt{u^2+f_1}}{f_1}.
\end{equation}
Let us consider Eq.~(\ref{georad}) in the limiting case $u\to0$.
Then,
\begin{equation}
\frac{du}{ds}\to-\frac{(r-r_-)(r-r_+)}{r^3f_1},
\label{georad2}
\end{equation}
where
\begin{equation}
r_\pm=\frac{m}{2\alpha}\left(\pm\sqrt{1+8\alpha}-1\right).
\label{rpmexpr}
\end{equation}
It can easily be seen that $r_-<0$ and $r_+<2m$ for $\alpha>0$;
therefore, $du/ds<0$ for $r>2m$. Since $u<0$, this means
that quantity $u$ remains negative and does not vanish
anywhere in region $r>2m$. Therefore, the deviation of
geodesics has no singularities at the boundaries of the
operation of coordinate systems $y\to y_1$ and $y\to y_2$ for
$2m<r<\infty$. Physically, this means that the tidal forces
acting on falling bodies are finite, and singularities for
$y=y_1$ and $y=y_2$ are of purely coordinate origin. These
singularities can be classified as the violation of metric
analyticity, which is not associated with the divergence
of algebraic invariants of the curvature tensor \cite{BroRub08}.

In the case of the emergent beam, additional singularities
appear for $y=y_3$ and $y=y_4$; these singularities will be
considered in Section~\ref{etaycoordsec} using other coordinates.


   \subsection{Accretion for $\alpha\geq1/8$}
   
Let us now suppose that $\alpha>1/8$ 
In this case, $y_1$ and $y_2$ are complex-conjugate
numbers. Evaluating the integral in expression (\ref{intsol}),
we obtain
\begin{eqnarray}
\label{Phi18}
\Psi&=&\frac{2}{\sqrt{(2y-1)^2+8\alpha-1}}\times
\\
&\times&\exp\left[-\frac{1}{\sqrt{8\alpha-1}}\arctan\frac{2y-1}{\sqrt{8\alpha-1}}\right].
\end{eqnarray}
Function $\Psi(y)$ is multivalued due to the arctangent,
but it does not affect the result because the twiddle
factor in expression (\ref{Phi18}) can be compensated for by
the choice of function $B(t)$, i.e., by the appropriate
redefinition of the time coordinate. Applying the same
method as in Section~\ref{diagtranssub}, we obtain
\begin{equation}
f_0=\frac{r^4[(2y-1)^2+8\alpha-1]^2}{16yr_0^4(2\alpha)^2\Phi^2}.
\end{equation}

In the intermediate case of $8\alpha=1$, we observe the
coincidence of two surfaces $y_1=y_2=1/2$ and
\begin{equation}
\Psi=\frac{1}{\left|y-\frac{1}{2}\right|}e^{1/(2y-1)},
\end{equation}
\begin{equation}
f_0=\frac{16r^4(y-1/2)^6}{yr_0^4}e^{-2/(2y-1)}.
\end{equation}
In a certain sense, this case is an analog of an extreme
black hole with two coinciding horizons.

We will not analyze the global geometry of the
resultant solutions in the $(t,r)$ coordinates in detail
because this can be done much more easily and effectively
in other coordinates $(\eta,y)$, which will be introduced
in Section~\ref{etaycoordsec}, where the Carter-Penrose diagrams
for all accretion cases will also be constructed.
In the $(\eta,y)$ coordinates, the analytic solution will be
obtained in explicit form, while the solution in the
$(t,r)$ coordinates can only be obtained in parametric
form.


   \subsection{Behavior of Divergence for $f_0\to\infty$ as $y\to 0$}
   
An interesting feature is the divergence $f_0\to\infty$ at
the visibility horizon of a Vaidya black hole for $y=0$. This behavior is opposite to that at the gravitation
radius of a Schwarzschild black hole, where $f_0=0$.
Accretion considerably changes the geometry near the
visibility horizon on account of the nonstationary
nature of the black hole. In this section, we explain
why this occurs. As an example, let us consider the
case of weak accretion. Relation $T_0^0+T_1^1=0$ and
Eqs.~(\ref{G00}), (\ref{G11}) give
\begin{equation}
\nu'-\lambda'=\frac{4M_1}{r^2}e^{\lambda}.
\label{lamnueq}
\end{equation}
from which we have
\begin{equation}
\nu'=\frac{f_0'}{f_0}=\frac{2M'_1/r+2M_1/r^2}{1-2M_1/r}.\label{m0f0m1f1}
\end{equation}
Integrating the latter relation, we obtain
\begin{equation}
f_0=e^{\phi(t)}\exp\left\{\int\limits_{r_0}^{r}d\tilde r\frac{2M'_1/\tilde r+2M_1/\tilde r^2}{1-2M_1/\tilde r}\right\},\label{f0phi}
\end{equation}
where $\phi(t)$ is an arbitrary function that can be excluded
by redefining time $dt'=dte^{\phi/2}$, and the integral in
Eq.~(\ref{f0phi}) is evaluated for $t=const$. The dependence on
$t$ and $\tilde r$ also appears in terms of variable $\tilde v(t,\tilde r)$ in function
$M_1(\tilde v)$, but the integral is evaluated along the line
$t=const$, i.e., time is a parameter in the integral.

The integral in the exponent in expression (\ref{f0phi}) can
be expressed in the form
\begin{eqnarray}
\nu=\int\limits_{r_0}^{r}&&d\tilde r\frac{2M'_1/\tilde r+2M_1/\tilde r^2}{1-2M_1/\tilde r}
\label{inttransf}
\\
&=&-\ln r^2-\ln\left|1-\frac{2M_1}{r}\right|+2\int\limits_{r_0}^{r}\frac{d\tilde r}{r\left(1-2M_1/\tilde r\right)},
\nonumber
\end{eqnarray}
where, as before, the integral is evaluated for $t=const$.
For $M_1=const$, the latter integral on the right-hand
side of expression (\ref{inttransf}) can be evaluated easily, and
expression (\ref{f0phi}) implies that, in this case, the
Schwarzschild metric with $f_0=f_1$ is realized.

Let us now consider the case when $\partial M_1/\partial v\neq0$,
i.e., with accretion $\partial M_1/\partial v>0$ in our situation. For
$t=const$, we can write $d\tilde r=f_1dv$. Then, the integral on
the right-hand side of Eq.~(\ref{inttransf}) can be written in the
following simple form:
\begin{equation}
\int\limits_{v_0}^{v}\frac{dv}{r(t,v)},\label{simp}
\end{equation}
where $v_0=const$. If we consider the motion of a
specific photon, we have $|v|=const<\infty$, and integral
(\ref{simp}) is bounded.

Let us trace the transition from the metric with
accretion of photons, for which $f_0\to\infty$ on the visibility
horizon, to the Schwarzschild metric with $f_0\to0$ at
the gravitational radius. The reason for such a radical
transformation is that variable $v$ at the end of accretion
(in the range of values of $\partial M_1/\partial v\to0$) tends to $-\infty$.
As a result, the integral in expression (\ref{inttransf}) tends to
$-\infty$, which can be seen from (\ref{simp}). This integral
becomes numerically equal to the second logarithm in
expression (\ref{inttransf}). For this reason, two infinities are cancelled
out at the visibility horizon, and ultimately $f_0\to0$ in the Schwarzschild metric. Thus, we can state that
the Schwarzschild metric is a degenerate special case.

Let us find the upper limit in expression (\ref{simp}). It is
determined by coordinate transformation (\ref{dudtdr}) at$t=const$, i.e., by the equation $rdr=[r-2M_1(v)]dv$. This
equation determines the section of surface$v=v(t,r)$
by plane $t=const$. For $M_1=const$, we obtain
\begin{equation}
v=r-2M_1+2M_1\ln|r-2M_1|+B_1(t)\to-\infty
\label{funvsch}
\end{equation}
for $r\to2M_1$ as mentioned above. For the
Schwarzschild metric, we have $B_1(t)=t+B_2$, where
$B_2=const$.


\section{COORDINATES $(\eta,y)$}
\label{etaycoordsec}

For a detailed analysis of the global geometry of the
Vaidya metric with a linear mass function $m(z)$, it is
expedient to transform Vaidya metric (\ref{Vaidya}) to the
orthogonal system with certain new coordinates $\eta$ and
$y$ (see \cite{we} for details) as follows:
\begin{equation}
ds^2=f_0(\eta,y) d\eta^2-\frac{dy^2}{f_1(\eta,y)}-r^2\left(d\theta^2+\sin^2\theta d\phi^2\right),
\label{newc}
\end{equation}
with two metric functions $f_0(\eta,y)$ and $f_1(\eta,y)$. The
first new variable $\eta$ will be defined later; for the second
variable, we choose $y=1-2m(z)/r$. For $m(z)=m_0-\alpha z$, calculations analogous to those in Section~\ref{diagtranssub} yield
\begin{equation}
m=C(\eta)\Phi(y),
\end{equation}
where
\begin{equation}
\label{Phi}
\Phi(y)\equiv\exp\left[-2\alpha\int\frac{dy}{(1-y)(y^2-y+4\alpha)}\right]>0,
\end{equation}
and $C(\eta)$ is an arbitrary function,
\begin{equation}
\label{f12}
f_1=-\frac{(1-y)^3(y^2-y+4\alpha)}{(2C\Phi)^2},
\end{equation}
\begin{equation}
\label{f02}
f_0=-\frac{(y^2-y+4\alpha)}{1-y}\frac{C_{,\eta}^2}{\alpha^2}\Phi^2.
\end{equation}
The roots of the equation $y^2-y+2\alpha=0$ were written
in (\ref{y12}), while the roots of the equation $y^2-y+4\alpha=0$
are given by expressions (\ref{z12}). Note that $0<y_1<y_3\leq1/2\leq y_4<y_2<1$.

By redefining time variable $\eta$, we can always make
$C_{,\eta}=const$. Then, the continuity of the limiting transition
to the Schwarzschild metric for $\alpha\to0$ always
requires that $C(\eta)=\alpha\eta+C_0$ with $C_0=const$.

\begin{figure}[h]
		\includegraphics[angle=0,width=0.5\textwidth]{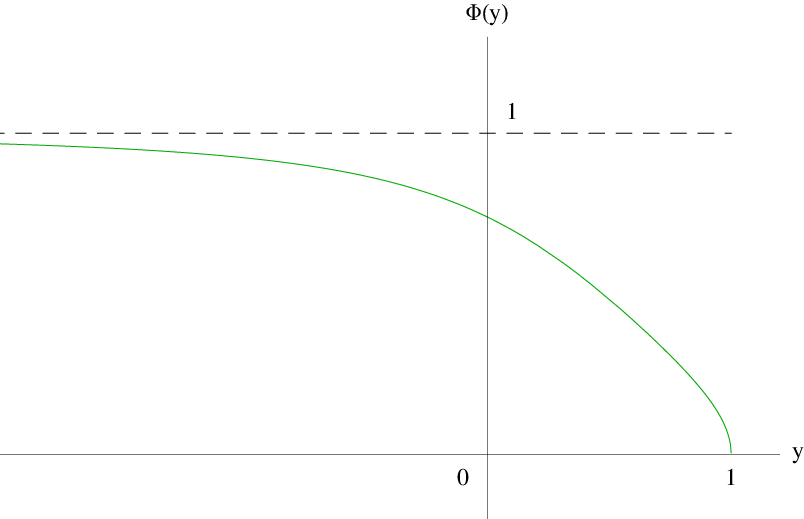}
		\includegraphics[angle=0,width=0.5\textwidth]{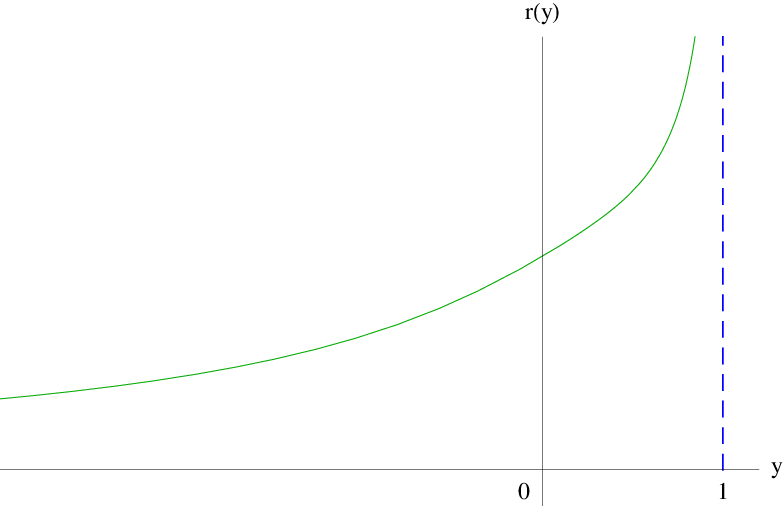}
	\caption{(upper panel) Function $\Phi(y)$ for $\alpha>1/16$. (lower panel) Function $r(y)$ for $\alpha>1/8$.}
	\label{Phigtr}
\end{figure}

As a result, we have determined the explicit dependence
of all metric functions in expression (\ref{newc}) on
coordinates $\eta$ and $y$. In evaluating the integral in
Eq.~(\ref{Phi}), we encounter three significantly different
situations, i.e., (i) powerful accretion for $\alpha>1/16$,
(ii) moderate accretion for $\alpha=1/16$, and (iii) weak
accretion for $\alpha<1/16$.

\begin{figure}[h]		
		\includegraphics[angle=0,width=0.5\textwidth]{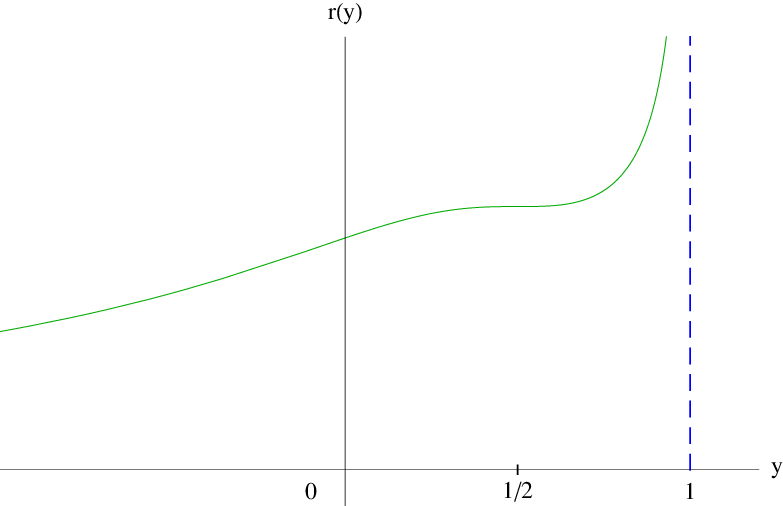}
        \includegraphics[angle=0,width=0.5\textwidth]{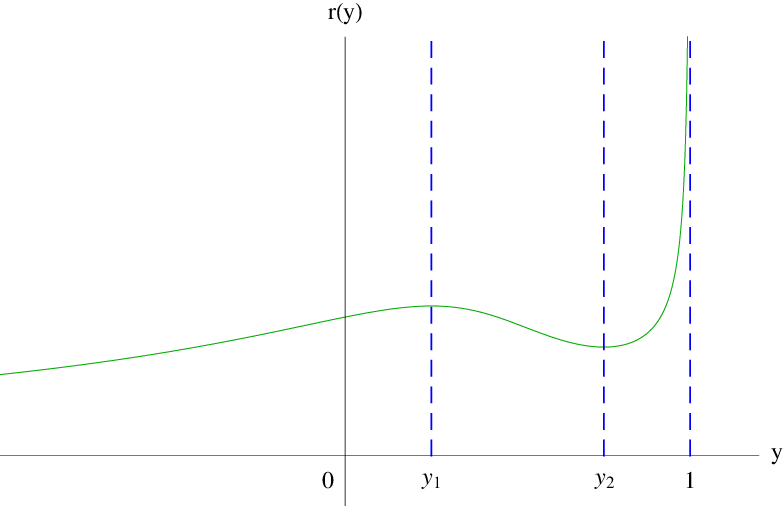}		
	\caption{(upper panel) Function $r(y)$ for $\alpha=1/8$. (lower panel) Function $r(y)$ in the case of intense accretion with $1/16<\alpha<1/8$.}
	\label{rgtr18}
\end{figure}

\begin{figure}[h]
	\begin{center}
		\includegraphics[angle=0,width=0.49\textwidth]{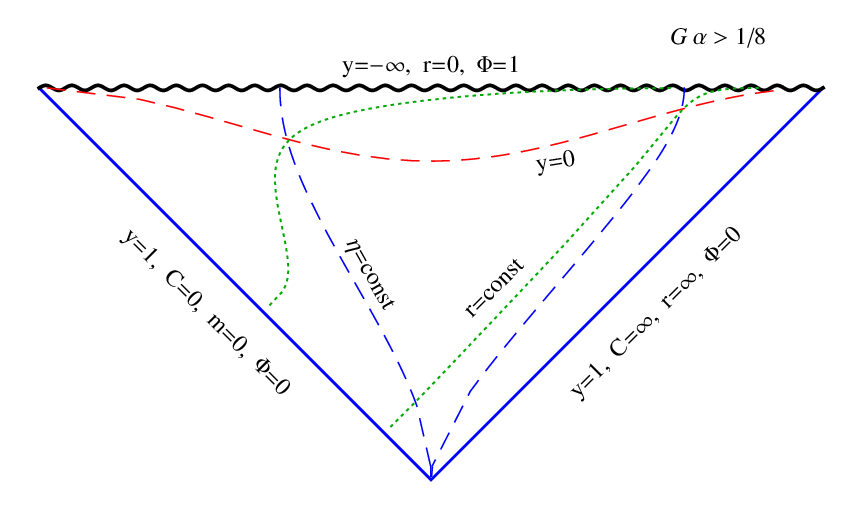}
	\end{center}
	\caption{Carter-Penrose diagram for global geometry of the
Vaidya metric in the case of superpower accretion with $\alpha>1/8$ with a linear mass function. Wavy line corresponds to
singularity $y=-\infty$ for $r=0$. One of two zeroth lines $y=1$
is zeroth infinity of past with $r=\infty$, while the other corresponds
to the boundary of the initial accretion with zero
mass ($m=0$). Time coordinate is measured from below,
and space coordinate is measured from left to right. In
region $T^*$, the time coordinate is ($-y$) everywhere. These
lines intersect spacelike line $y=0$, which is the visibility
horizon and separates the spacelike region, in which surfaces
$r=const$ are timelike ($y>0$), from the regions in
which these surfaces are spacelike ($y<0$).}
	\label{diagr-superpower}
\end{figure}

Let us consider invariant (\ref{Y0}) in the $(\eta,y)$ coordinates.
We will refer to a region of space-time as the $R^*$
region in which $Y<0$ and as the $T^*$ region a region in
which $Y>0$. In the $R^*$ regions, $\eta$ is the time coordinate
and $y$ is the space coordinate; conversely, $\eta$ is the
space coordinate in $T^*$ regions and $y$ is the time coordinate
in $T^*$ regions.

The metric assumes the following simple form after
the conformal transformation:
\begin{eqnarray}
\label{conformalmetric}
ds^2&=&\frac{C^2\Phi^2}{\alpha^2(1-y)}\left\{(y^2-2y+4\alpha)
\left[(d\log\Phi)^2)-(d\log C)^2\right]\right\}-
\nonumber
\\
&-&r^2d\Omega^2.
\end{eqnarray}
It can be seen that zero geodesics are defined by the
equations
\begin{equation}
\label{redialnull}
C=A\, \Phi^{\pm1}, \quad A=const.
\end{equation}
In the case of accretion, the upper superscript ``+''
corresponds to emergent beams and the lower subscript
``-'' corresponds to incident zeroth beams, and
vice versa in the case of the Vaidya metric formed by
emergent radiation.

Taking into account expression (\ref{conformalmetric}), we construct
the Carter-Penrose diagrams in the $\log C$ vs. $\log\Phi(y)$
coordinates using the following transformation:
\begin{eqnarray}
t'&=&\arctan\left[\log C + \log\Phi(y)\right] - \arctan\left[\log C - \log\Phi(y)\right]\nonumber
\\
x'&=&\arctan\left[\log C  + \log\Phi(y)\right] + \arctan\left[\log C - \log \Phi(y)\right],\nonumber
\end{eqnarray}
with corresponding shifts and variable of the axes
whenever required. 

\begin{figure}[h]
	\begin{center}
		\includegraphics[angle=0,width=0.49\textwidth]{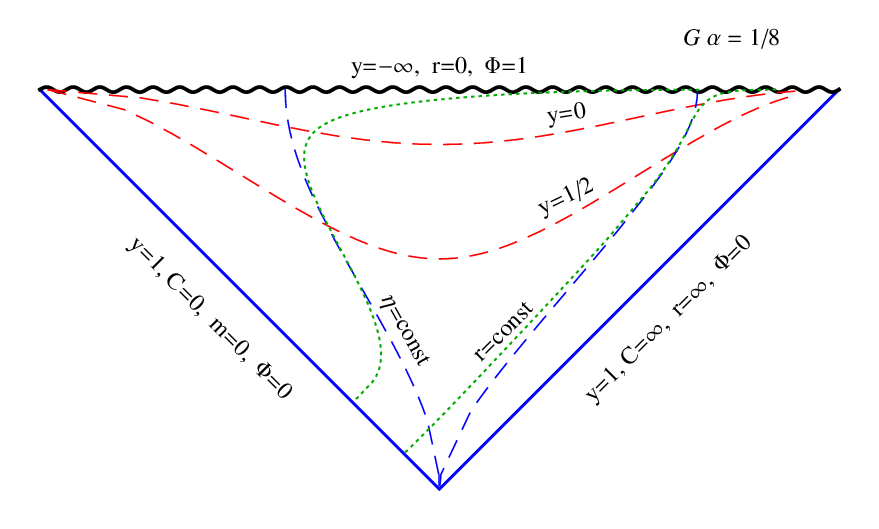}
	\end{center}
	\caption{Carter-Penrose diagram for the case of transition
from the superpower to simply powerful accretion with $\alpha=1/8$. This case differs from the previous one in the existence
of turning points on lines$r=const$ corresponding to
$y=y_1=y_2=1/2$ on lines $r=r(y)$.}
	\label{diagr-superpower18}
\end{figure}

Let us begin with the case of high-power accretion
($\alpha>1/16$). Then, $Y>0$ and, hence, we are in the $T^*$
region. Integration in Eq.~(\ref{Phi}) yields
\begin{eqnarray}
\Phi&=&\frac{\sqrt{1-y}}{(y^2-y+4\alpha)^{1/4}}\times
\nonumber
\\
&\times&\exp\left[-\frac{1}{2\sqrt{16\alpha-1}}\left(\arctan\frac{2y-1}{\sqrt{16\alpha-1}}+\frac{\pi}{2}\right)\right].
\label{PhiLarge}
\end{eqnarray}

The behavior of function $\Phi$ is qualitatively the
same in the entire region $\alpha>1/16$ (see Fig.~\ref{Phigtr}). However,
dependence $r(y)$ behaves differently in the intervals
$1/16<1/8<\alpha$, $\alpha=1/8$, and $1/16<\alpha<1/8$ (see Figs.~\ref{Phigtr},~\ref{rgtr18}). In accordance with expression (\ref{PhiLarge}), radius$r=2m/(1-y)$ for $\alpha>1/8$ is a monotonically increasing
function of $y$ (from $r=0$ for $y=\infty$ to $r=\infty$ for $y=1$).
We will refer to the case with $\alpha>1/8$ as superpower
accretion and the case with $1/16<\alpha<1/8$ as simply
strong accretion. It should also be noted that curves
$y=y_1$ and $y=y_2$ are spacelike curves.

Apart from natural boundaries such as $r=0$ and
infinities, the Carter-Penrose diagrams also display
horizons of different types (zeroth, timelike, and
spacelike), which represent the boundaries of diagrams;
in this case, spherically symmetric spacetime
consists of a certain set of triangles and squares separated
by common boundaries. It should be noted that
we have imposed additional physical requirement
$m\geq0$. As a result, physical spacetime may turn out to
be geodetically incomplete.

Since $\Phi(y)>0$ by definition, our physical limitation
$m\geq0$ leads to inequalities $C\geq0$, $y\leq1$, while $r\geq0$
implies that $-\infty<y<1$. Therefore, the boundaries are
$y=-\infty$ and $y=1$. Everywhere, we have $Y\geq0$, i.e., the
$T^*$- region lies within the boundaries, where lines $y=const$ are spacelike ($(-y)$ plays the role of time, and this
time increases from below), while the lines $\eta=const$
(or $C(\eta)=const$) are timelike. Then, we see that
boundaries $y=-\infty$ ($r=0$) are spacelike and singular
because invariant $Y\to+\infty$ for $y\to1$.

\begin{figure}[h]
	\begin{center}
		\includegraphics[angle=0,width=0.49\textwidth]{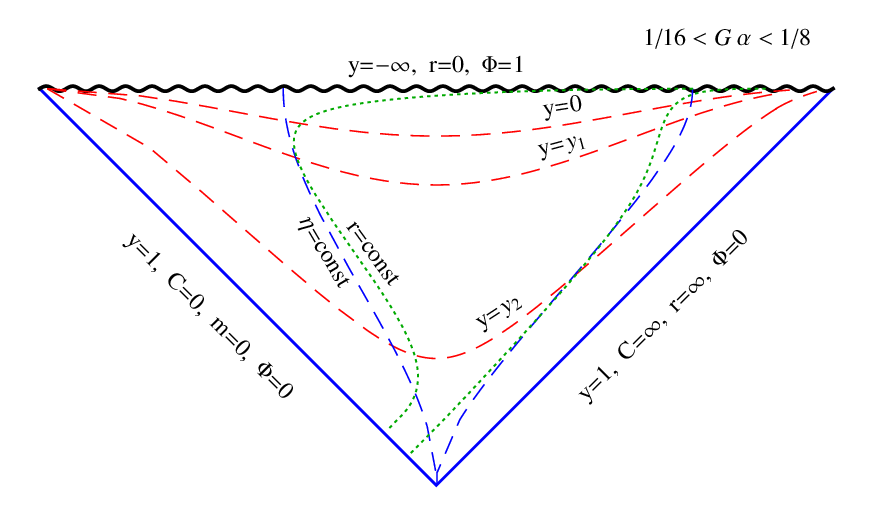}
	\end{center}
	\caption{Carter-Penrose diagram for simply powerful accretion
with $1/16<\alpha<1/8$.}
	\label{diagr-power}
\end{figure}

\begin{figure}[h]
	\includegraphics[angle=0,width=0.5\textwidth]{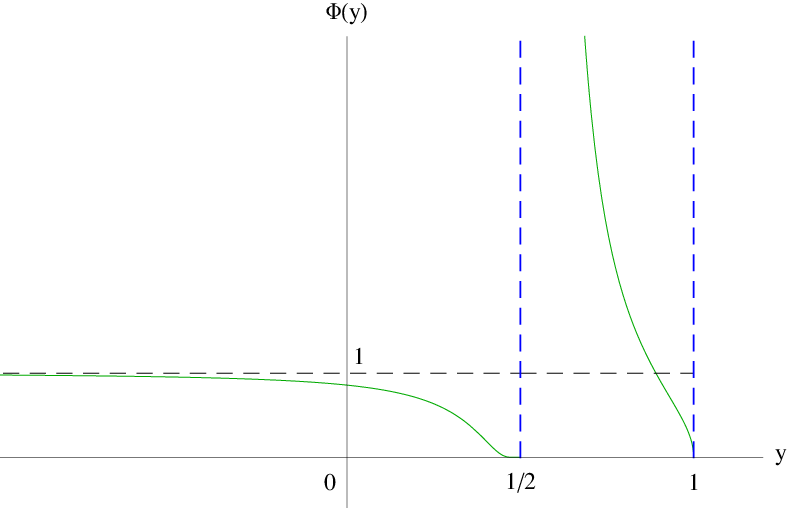}
	\includegraphics[angle=0,width=0.5\textwidth]{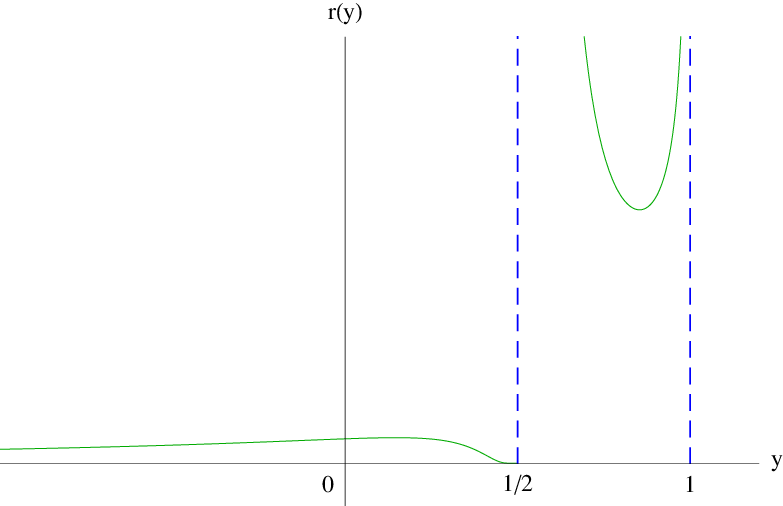}
	\caption{Functions (upper panel) $\Phi(y)$ and (lower panel) $r(y)$) for $\alpha=1/16$.}
	\label{phi-superpower116}
\end{figure}

\begin{figure}[h]
\begin{center}
		\includegraphics[angle=0,width=0.49\textwidth]{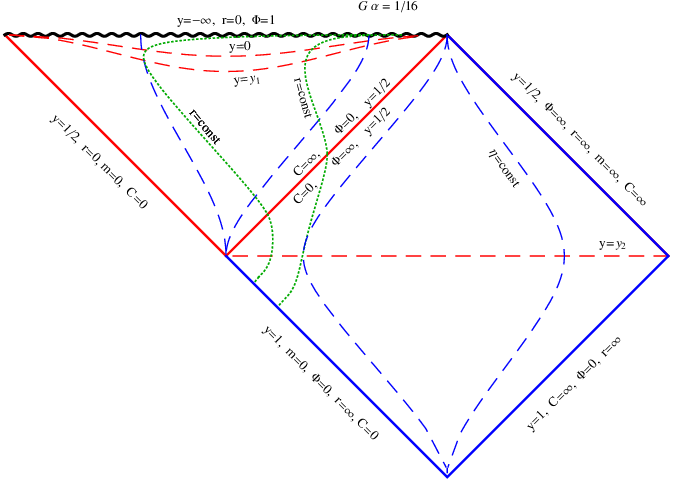}
\end{center}		
		\caption{Carter-Penrose diagram for the case of transition from powerful to weak accretions with $\alpha=1/16$. Double horizon $y=y_3=y_4=1/2$ separates two $T^*$ regions. Zeroth infinity of future appears, where $y=1/2$ and $m$, $r=\infty$.}
		\label{diagr-superpower116}
\end{figure}

At first glance, the boundaries for $y=1$ are zeroth
boundaries. However, since $C(\eta)$ and $\Phi(y)$ are in the
denominator of $Y$ and can vanish or turn to infinity, a
more meticulous analysis is required. We will use two
congruences of zeroth geodesics. Let us first consider
incident beams for which $m=C(\eta)\Phi(y)=const$.
These beams begin from $y=1$, where $r\propto1/(1-y)$, $Y\propto(1-y)^3$, and enter the spacetime singularity $r=0$.
In can be seen that the boundary is indeed the zeroth
infinity of past, where $y=1$ and $r=\infty$. The value of $m$
along this boundary varies from $m=0$ to $m=\infty$, while
$\Phi(y)=0$ and $C(\eta)=\infty$.

\begin{figure}[h]
	\includegraphics[angle=0,width=0.5\textwidth]{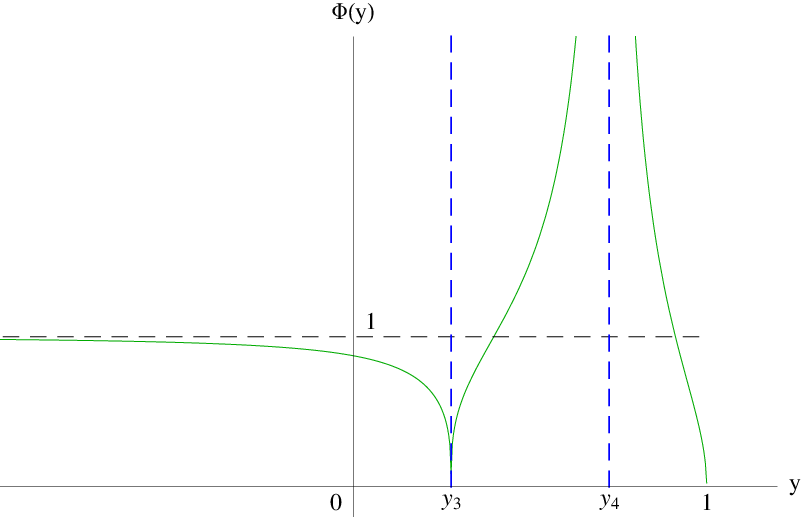}
	\includegraphics[angle=0,width=0.5\textwidth]{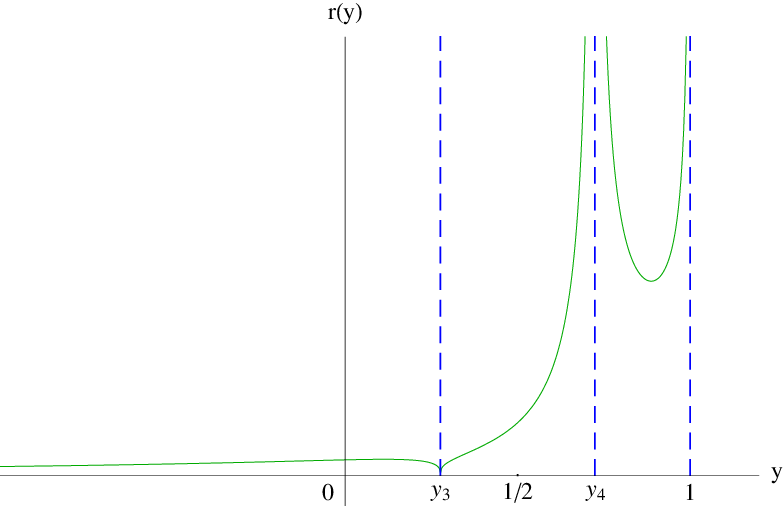}
	\caption{Functions (upper panel) $\Phi(y)$ and (lower panel) $r(y)$ for weak accretion with $\alpha<1/16$.}
	\label{phi-slow}
\end{figure}

The origin of the second boundary $y=1$ is more
complicated. Let us consider the second congruence
of zeroth geodesics for which $C(\eta)=const\cdot \Phi(y)$. They
begin from $y=1$, where $m\propto(1-y)$, $r=const$, and
$Y\propto(1-y)$ for $y\to1$. Therefore, the boundary $y=1$
under investigation is a zeroth boundary, along which $\Phi(y)=0$, $C(\eta)=0$, $m=0$, and $r$ varies from $r=0$ to
$r=\infty$. Consequently, it is no longer infinite, but is the
edge of zeroth beams that initiate accretion. This
spacetime is obviously not geodetically complete (see
Fig.~\ref{diagr-superpower}). Analogous arguments for $\alpha=1/8$ and $1/16<\alpha<1/8$ lead to the diagrams shown in Figs.~\ref{diagr-superpower18} and \ref{diagr-power}.

In the intermediate case of $\alpha=1/16$, function $\Phi(y)$
has the form
\begin{equation}
\label{Phi2}
\Phi=\sqrt{\left|\frac{y-1}{y-(1/2)}\right|}\exp\left\{\frac{1}{4[y-(1/2)]}\right\},
\end{equation}
and
\begin{equation}
f_0=\left|y-\frac{1}{2}\right|\exp\left\{\frac{1}{2\left(y-\frac{1}{2}\right)}\right\},
\end{equation}
\begin{eqnarray}
Y&=&\frac{(1-y)^3(y-\frac{1}{2})^2}{4C^2\Phi^2}=\nonumber
\\
&=&\frac{(1-y)^2\left|y-\frac{1}{2}\right|^3}{4C^2}\exp\left\{-\frac{1}{2\left(y-\frac{1}{2}\right)}\right\},
\end{eqnarray}
(see Fig.~\ref{phi-superpower116}). The Carter-Penrose diagram consists of
two parts, viz., a triangle and a square joined together
(and separated) by double horizon $y=y_3=y_4=1/2$.
Regions $T^*$ lie on both sides of this horizon, and lines
$y=const$ are spacelike.

The triangle consists of spacelike singular lines$y=-\infty$, where $r=0$, and two zeroth boundaries $y=y_3=y_4=1/2$. One of these boundaries is a double horizon,
while the other is a boundary zeroth beam with $m=0$
and $r=0$. Operations with double horizons require
carefulness because $\Phi(1/2-0)=0$ and $\Phi(1/2-0)=\infty$. Let us consider the beams with $C\Phi=m=const$,
which produce accretion. It can be seen that $C(\eta)=\infty$
for $y(1/2-0)$, and it is just the end of the range of
coordinates in the triangle. For $y(1/2+0)$, we have
$C(\eta)=0$, and it is the beginning of the new spatial
range of coordinates in the square. For the boundary
beam with $m=0$, both $C$ and $\Phi$ are equal to zero along
$y=1/2$. Invariant $Y$ diverges, but point $y=1/2$ is a
coordinate singularity (the determinant of the metric
tensor is equal to zero at this point; see the description
of the construction in Fig.~\ref{diagr-superpower116}).

In the case of weak accretion ($\alpha<1/16$), the double
horizon splits into two horizons for $y=y_3$ and $y=y_4$. The new horizons appear due to the fact that, here,
we are dealing with unbounded accretion with an
infinite increase in the black hole mass. Function $\Phi(y)$
now assumes the form
\begin{equation}
\label{Phiweak}
\Phi=\sqrt{1-y}\,|y-y_3|^{y_3/[2(y_4-y_3)]}\,|y-y_4|^{-y_4/[2(y_4-y_3)]},
\end{equation}
and
\begin{equation}
f_0=-|y-y_3|^{\frac{y_3}{y_4-y_3}+1}|y-y_4|^{\frac{-y_4}{y_4-y_3}+1},
\end{equation}
\begin{equation}
Y=\frac{(1-y)^2(y-y_3)(y-y_4)}{4C^2}|y-y_3|^{\frac{-y_3}{y_4-y_3}}|y-y_4|^{\frac{y_4}{y_4-y_3}},
\end{equation}
(see the graphs in Fig.~\ref{phi-slow}).

\begin{widetext}
\begin{figure}[t]
\begin{center}
		\includegraphics[angle=0,width=0.8\textwidth]{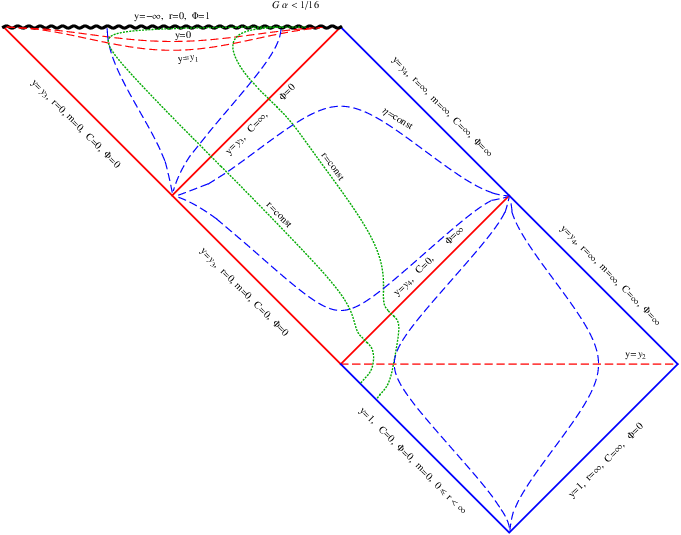}
\end{center}
		\caption{Carter-Penrose diagram for the case of weak accretions with $\alpha<1/16$. Double horizon splits into two parts: event horizon
for $y=y_3$ and cosmological horizon for $y=y_4$. Regions $T^*$ now exist for $-\infty<y<y_3$ and for $y_4<y<1$. The $R^*$ region in
which $\eta$ is the time coordinate and $y$ is the space coordinate lies in the interval $y_3<y<y_4$.}
		\label{diagr-slow}
\end{figure}
\end{widetext}

In this case, the global geometry is more complicated.
The Carter-Penrose diagram consists of one
triangle and two squares. The structure of the boundaries
of the triangles remains the same as before (but
now $y<1/2$), and we have the $T^*$ region in which the
line $\eta=const$  is timelike, while $y=const$ is spacelike.
The $R^*$ region also appears, the boundaries of which
are new horizons $y=y_3$ and $y=y_4$. The left boundary
consists of two parts for $y=y_3$. The latter is just the
extreme zeroth beam of accretion with $m=0$, $C(\eta)=0$, $\Phi=0$, and $r=0$, while the upper boundary with
$C=\infty$ and $\Phi=0$ is an event horizon between the $R^*$
region and the $T^*$ region. The right boundary also
consists of two parts for $y=y_4$. The upper part of the
boundary is the last accretion beam with $r=\infty$, $m=\infty$, $C(\eta)=\infty$, and $\Phi=\infty$ (zeroth infinity of
future), while the lower boundary is a cosmological
horizon connecting the $R^*$ region and the outer $T^*$
region ($y_3\leq y_4\leq1$). The second square ($y_4\leq y\leq1$) has
the same structure except for the fact that now $y_4>1/2$. As mentioned above, such spacetime is not geodetically
complete (see the corresponding diagram in
Fig.~\ref{diagr-slow}).

\section{CONCLUSIONS}
\label{concl}

In this study, we have obtained the transformation
of coordinates from the standard representation of the
Vaidya metric with a linear mass function to two diagonal
systems of coordinates $(t,r)$ and $(\eta,y)$. The
advantage of the linear model under investigation lies
in the possibility of analytic calculation of all metric
functions and light geodesics. It turns out that, in the
presence of even weak accretion near the horizon,
there is a narrow region in which the solution differs
from the Schwarzschild solution not only quantitatively,
but also qualitatively. Namely, apart from the
visibility horizon, there are surfaces in diagonal coordinates
with metric singularities $g_{00}\to 0$ and $g_{00}\to\infty$,
which are surfaces of the infinite red and blue shifts,
respectively. These surfaces serve as the boundaries of
various coordinate systems, and their appearance distinguishes
qualitatively the resultant metric of an
accreting black hole from the Schwarzschild metric. It
has been shown that one coordinate system with diagonal
coordinates is insufficient for covering the entire
spacetime; several such systems are required. The difference
from the Schwarzschild metric (for example,
in the case of very weak accretion) is associated with
the divergence of coordinate time t for a radial light
beam incident on the surface located outside the visibility
horizon of a Vaidya black hole. In this case, the
coordinate time of radially propagating photons on the
visibility horizon turns out to be finite.

The divergence of the energy-momentum tensor
components $T^{00}$ and $T^{11}$ on these surfaces is not associated
with the presence of physical caustic, but is of
purely coordinate origin. Indeed, analysis of the deviation
equations for geodesics has shown that tidal forces are finite on surfaces $g_{00}\to 0$ and $g_{00}\to\infty$ in the
$R$ region (for $r>2m$). For this reason, these surfaces
are exclusively coordinate singularities, which can be
referred to as false firewalls.

In the second set of diagonal coordinates $(\eta,y)$, we
have determined the maximal analytic continuation of
the Vaidya metric in various cases that correspond to
different accretion rates and have managed to construct
a complete set of Carter-Penrose diagrams.
These diagrams contain a set of spacetime regions separated
by horizons and boundary lines (either $g_{00}=0$
or $g_{00}=\infty$), on which the operation of the coordinate
systems terminates. The spacetime on the constructed
diagrams is geodetically incomplete because we have
imposed the physical condition of nonnegativity of
mass function ($m\geq0$); however, this construction is
the maximum possible in the given physical formulation
of the problem.

This study was supported by the Russian Foundation
for Basic Research (project no. 15-02-05038-a).

\end{document}